\documentclass[a4paper,11pt]{article}
\usepackage{ifluatex}
\ifluatex 
\usepackage[utf8]{luainputenc}
\else
\usepackage[utf8]{inputenc}
\fi

\usepackage{jinstpub} 

\usepackage{lineno}

 \usepackage{tikz}
\usepackage{tikz-imagelabels}
\usetikzlibrary{calc,arrows, arrows.meta,decorations.pathreplacing}

\usepackage{graphicx}
\usepackage{siunitx}
\usepackage{setspace}
\usepackage{float}
\usepackage{caption,subcaption}
\usepackage{media9}
\usepackage{multimedia}
\usepackage{tabularx}
\usepackage{wrapfig}
\usepackage{comment}
\usepackage{calc}
\usepackage{multicol}

\title{\boldmath Method and portable bench for tests of the laser optical calibration system components for the Baikal-GVD underwater neutrino Cherenkov telescope}

\author[a]{V.A.~Allakhverdyan}
\author[b]{A.D.~Avrorin}
\author[b]{A.V.~Avrorin}
\author[b]{V.M.~Aynutdinov}
\author[c]{R.~Bannasch}
\author[d]{Z.~Barda\v{c}ov\'{a}}
\author[a]{I.A.~Belolaptikov}
\author[a]{I.V.~Borina}
\author[a]{V.B.~Brudanin}
\author[e]{N.M.~Budnev}
\author[a]{V.Y.~Dik}
\author[b]{G.V.~Domogatsky}
\author[b]{A.A.~Doroshenko}
\author[a,d]{R.~Dvornick\'{y}}
\author[e]{A.N.~Dyachok}
\author[b]{Zh.-A.M.~Dzhilkibaev}
\author[d]{E.~Eckerov\'{a}}
\author[a]{T.V.~Elzhov}
\author[f]{L.~Fajt}
\author[g]{S.V.~Fialkovski}
\author[e]{A.R.~Gafarov}
\author[b]{K.V.~Golubkov}
\author[a]{N.S.~Gorshkov}
\author[e]{T.I.~Gress}
\author[a]{M.S.~Katulin}
\author[c]{K.G.~Kebkal}
\author[c]{O.G.~Kebkal}
\author[a]{E.V.~Khramov}
\author[a]{M.M.~Kolbin}
\author[a]{K.V.~Konischev}

\author[h,1]{K.A.~Kopa\'{n}ski, \note{Corresponding author.}}

\author[a]{A.V.~Korobchenko}
\author[b]{A.P.~Koshechkin}
\author[i]{V.A.~Kozhin}
\author[a]{M.V.~Kruglov}
\author[b]{M.K.~Kryukov}
\author[g]{V.F.~Kulepov}
\author[h]{Pa.~Malecki}
\author[a]{Y.M.~Malyshkin}
\author[b]{M.B.~Milenin}
\author[e]{R.R.~Mirgazov}
\author[a]{D.V.~Naumov}
\author[a]{V.~Nazari}
\author[h]{W.~Noga}
\author[b]{D.P.~Petukhov}
\author[a]{E.N.~Pliskovsky}
\author[j]{M.I.~Rozanov}
\author[a]{V.D.~Rushay}
\author[e]{E.V.~Ryabov}
\author[b]{G.B.~Safronov}
\author[a]{B.A.~Shaybonov}
\author[b]{M.D.~Shelepov}
\author[a,d,f]{F.~\v{S}imkovic}
\author[a]{A.E. Sirenko}
\author[i]{A.V.~Skurikhin}
\author[a]{A.G.~Solovjev}
\author[a]{M.N.~Sorokovikov}
\author[f]{I.~\v{S}tekl}
\author[b]{A.P.~Stromakov}
\author[a]{E.O.~Sushenok}
\author[b]{O.V.~Suvorova}
\author[e]{V.A.~Tabolenko}
\author[e]{B.A.~Tarashansky}
\author[a]{Y.V.~Yablokova}
\author[c]{S.A.~Yakovlev}
\author[b]{D.N.~Zaborov}

\affiliation[a]{Joint Institute for Nuclear Research, Dubna, Russia}
\affiliation[b]{Institute for Nuclear Research, Russian Academy of Sciences, Moscow, Russia}
\affiliation[c]{EvoLogics GmbH, Berlin, Germany}
\affiliation[d]{Comenius University, Bratislava, Slovakia}
\affiliation[e]{Irkutsk State University, Irkutsk, Russia}
\affiliation[f]{Czech Technical University in Prague, Prague, Czech Republic}
\affiliation[g]{Nizhny Novgorod State Technical University, Nizhny Novgorod, Russia}
\affiliation[h]{Institute of Nuclear Physics of Polish Academy of Sciences (IFJ~PAN), Krak\'{o}w, Poland}
\affiliation[i]{Skobeltsyn Institute of Nuclear Physics, Moscow State University, Moscow, Russia}
\affiliation[j]{St.~Petersburg State Marine Technical University, St.Petersburg, Russia}

\emailAdd{konrad.kopanski@ifj.edu.pl}

\abstract{Large-scale deep underwater Cherenkov neutrino telescopes like Baikal-GVD, ANTARES or KM3NeT, require calibration and testing methods of their optical modules. These methods usually include laser-based systems which allow us to check the telescope responses to the light and for real-time monitoring of the optical parameters of water such as absorption and scattering lengths, which show seasonal changes in natural reservoirs of water. We will present a testing method of a laser calibration system and a set of dedicated tools developed for Baikal- GVD, which includes a specially designed and built, compact, portable, and reconfigurable scanning station. This station is adapted to perform fast quality tests of the underwater laser sets just before their deployment in the telescope structure, even on ice, without a  darkroom. The testing procedure includes the energy stability test of the laser device, 3D scan of the light emission from the diffuser and attenuation test of the optical elements of the laser calibration system. The test bench consists primarily of an automatic mechanical scanner with a movable Si detector, beam splitter with a reference Si detector and, optionally, Q-switched diode-pumped solid-state laser used for laboratory scans of the diffusers. The presented test bench enables a 3D scan of the light emission from diffusers, which are designed to obtain the isotropic distribution of photons around the point of emission. The results of the measurement can be easily shown on a 3D plot immediately after the test and may be also implemented to a dedicated program simulating photons propagation in water, which allows us to check the quality of the diffuser in the scale of the Baikal-GVD telescope geometry.}


\collaboration[c]{on behalf of Baikal-GVD collaboration}

\begin{document}
\maketitle
\pagebreak
\setcounter{page}{1}
\section{Introduction}
\label{sec:intro}
\begin{figure}[H]
	\vspace*{-2ex}
\begin{minipage}{0.75\textwidth}
	\includegraphics[width=\linewidth,keepaspectratio]{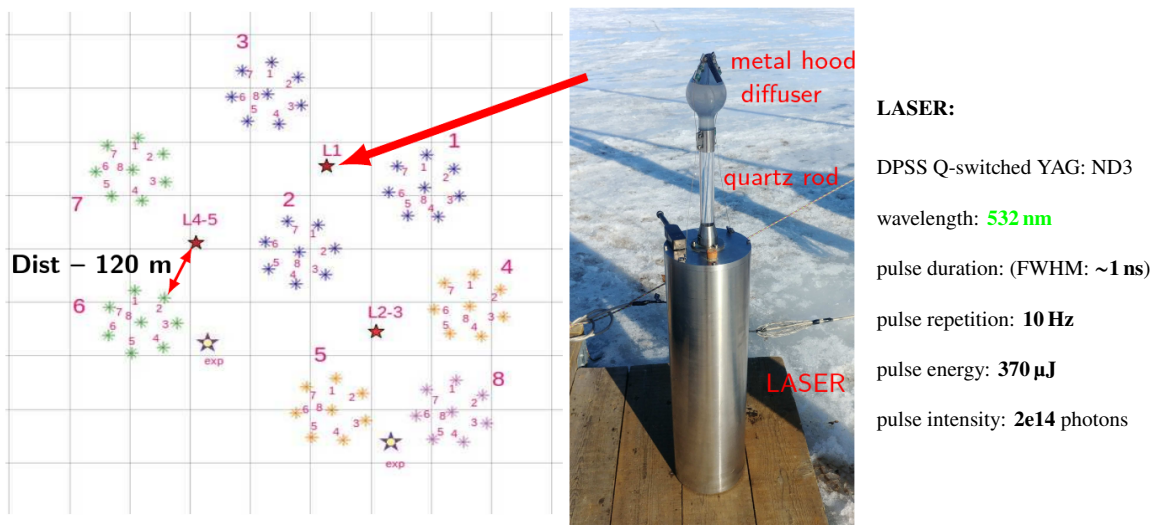}
\end{minipage}%
\hspace*{1ex}
\begin{minipage}{0.3\textwidth}
				
				\flushleft
				\scriptsize
				\setstretch{2.0}
				\textbf{LASER:}\\\vspace*{1ex}
				DPSS Q-switched YAG: ND3\\
				wavelength: {\color{green}\textbf{\SI{532}{\nano \meter}}}\\
				\mbox{pulse~duration:~(FWHM:~\textbf{\SI{\sim 1}{ns}})}\\
				pulse repetition: \textbf{\SI{10}{\hertz}}\\
				pulse energy: \textbf{\SI{370}{\micro \joule}}\\
				pulse intensity: \textbf{2e14} photons\\
\end{minipage}
\caption{Left: Clusters map for 2021. The positions of the lasers (L1-L5) are marked by red stars. Strings with lasers are placed about 120 m from the nearest instrumented string. Right: Laser in a pressure housing, equipped with a lightguide and a diffuser with a metal hood (reflector) on the top.}
\label{fig:laser_description}
\vspace*{-2ex}
\end{figure}

The Baikal-GVD telescope\cite{gvdcollaboration2021measuring} is located in the south-eastern part of Lake Baikal, about 55 km
to the south of Irkutsk. The telescope’s clusters are deployed in a distance around 3.6 km from
the shore at a depth of 1366 m (anchor). The telescope currently consists of eight clusters, which
are the independent telescope units. Each cluster consists
of 8 instrumented strings, on each of them there are 36 optical modules, grouped into sections
of 12 optical modules. A diagram of
the cluster distribution is shown in Fig. \ref{fig:laser_description}.
In such a large telescope, it is necessary
to synchronize the time as precisely as possible between individual modules. To ensure this, several
methods of optical calibration are used\cite{article_baikal}: LEDs placed in the optical modules, which by emitting
short ﬂashes of light enable calibration of the optical modules (intra-section callibration), LED
matrix used for inter-section calibration, and a laser system used for calibration between clusters.
Lasers (marked in the Fig. 1) are placed between the clusters of the telescope at a distance of about
120 m from the nearest instrumented string.
Currently, all lasers are equipped with optical diffusers, enabling a fairly isotropic distribution of
light around the laser device. For the proper operation of the entire system, a dedicated simulation
is also necessary, to determine both the amount of light registered by the modules, as well as
to precisely estimate the time of arrival of subsequent signals. Such simulation allows accurate determination of light propagation during a real event recorded by the telescope. 
To increase the accuracy of the calibration, a special test bench was developed, enabling not only precise examine of light distribution from diffusers
in laboratory conditions, but also to thoroughly examine the laser system components, including the lasers themselves, before their deployment from the surface of a frozen lake.
Such a measurement allows to obtain a data set required to estimate the aging rate of laser devices (lasers and optics as
well), but most importantly, it enables actual input data to be obtained for simulation. The method,
being the subject of this publication, is a combination of a hardware solution, which is the automatic,
reconfigurable test bench, with simulation of light propagation in water. This method is therefore
an extremely useful, efficient and accurate tool that allows for a quick determination of the impact
of laser system modification on the results observed in the structure of a real telescope. The
simulation program was called Pretorian, because it was created, inter alia, to ensure the safety of
the Optical Modules during in-situ tests with a laser beam distributed by other methods (collimated
beam, cone), than by using a diffuser.
%
\section{Test bench and method description}

\begin{figure}[H]
	\vspace*{-1.5ex}
	\begin{minipage}[t]{0.31\textwidth}\vspace{0pt}
		\includegraphics[width=\textwidth,keepaspectratio]{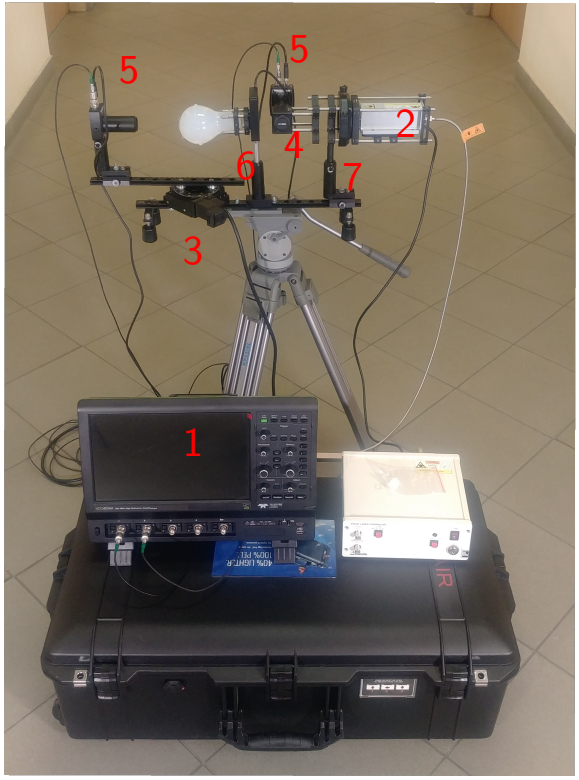}
	\end{minipage}\hfill
	\begin{minipage}[t]{0.62\textwidth}
		\caption{The scanner consists of a test bench (3) equipped with two automatic rotary tables -- one placed horizontally (rotation in the horizontal plane) and the other placed vertically (rotation of the tested element around its axis of symmetry in the axis of the laser beam). The test bench is also equipped with reconfigurable component holders (6), laser holder (7), beam splitter with attenuators set (4) and two PDA 100A2 detectors (5) -- reference and measurement. The entire test bench, together with the Q-switched DPSS laboratory laser (\SI{532}{\nano\meter}, \SI{137}{\micro \joule}) and the necessary accessories, is housed in a reinforced transportation box, which allows it to be easily moved and transported to perform the complex tests during a Baikal-GVD Winter Expedition.}
		\label{fig:slajd6}
	\end{minipage}
\vspace*{-4ex}
	
\end{figure}

Test bench (Fig. \ref{fig:slajd6}) is operated from the Teledyne LeCroy HDO 4034A oscilloscope (1). For this purpose, a special control program has been developed, which also performs data acquisition from the oscilloscope to a file. This allows us to easily use the obtained data set from the scan in the dedicated simulation. The program also makes a real time plot from each scanned plane.

After making a full scan, program creates a 3D graphic presenting the distribution of light intensity on the surface of the tested diffuser (option available in the diffuser test mode). Along with the test bench, the PM100USB laser beam energy monitor with the ES11C detector is also used to control the stability of the laser parameters. The stand can work with both a laboratory laser and the laser used in the calibration system of the Baikal-GVD telescope. Thanks to use of Q-switched lasers, it is not necessary to use a darkroom -- measurements can be made even in the sunlight. 

\begin{figure}[H]
	\vspace*{-2ex}
	\centering
	\begin{subfigure}[t]{\textwidth}
	\includegraphics[width=\textwidth,keepaspectratio]{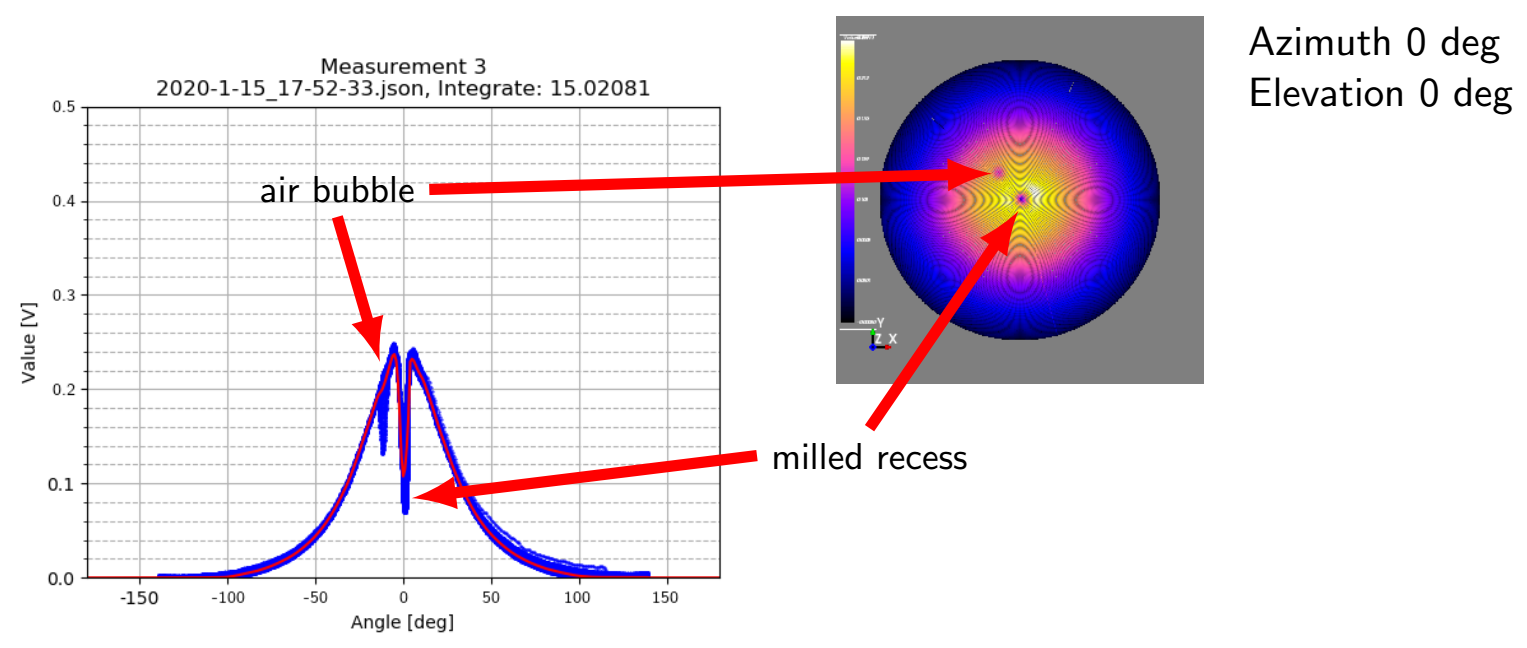}
	\end{subfigure}
	\hfill
	\begin{subfigure}[t]{\textwidth}
		\vspace*{-17ex}
	\hfill
	\includegraphics[width=0.25\textwidth,keepaspectratio]{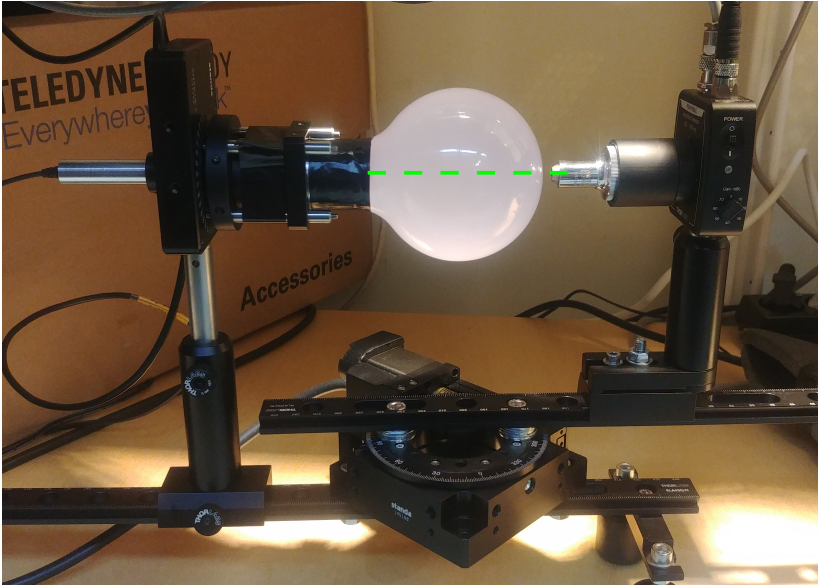}
	\end{subfigure}
	\vspace*{-5ex}
	\caption{Near scan result. The plot shows averaged measurements from many planes. On the right side the 3D visualization showing the upper part of the diffuser. On the picture below, the configuration of the test bench for near scan. The dashed green line represents the Z axis, perpendicular to the picture above.}
	\label{fig:near_scan}
	\vspace*{-3ex}
\end{figure}

\begin{figure}[H]
	\vspace*{-2ex}
	\centering
	\includegraphics[width=\textwidth,keepaspectratio]{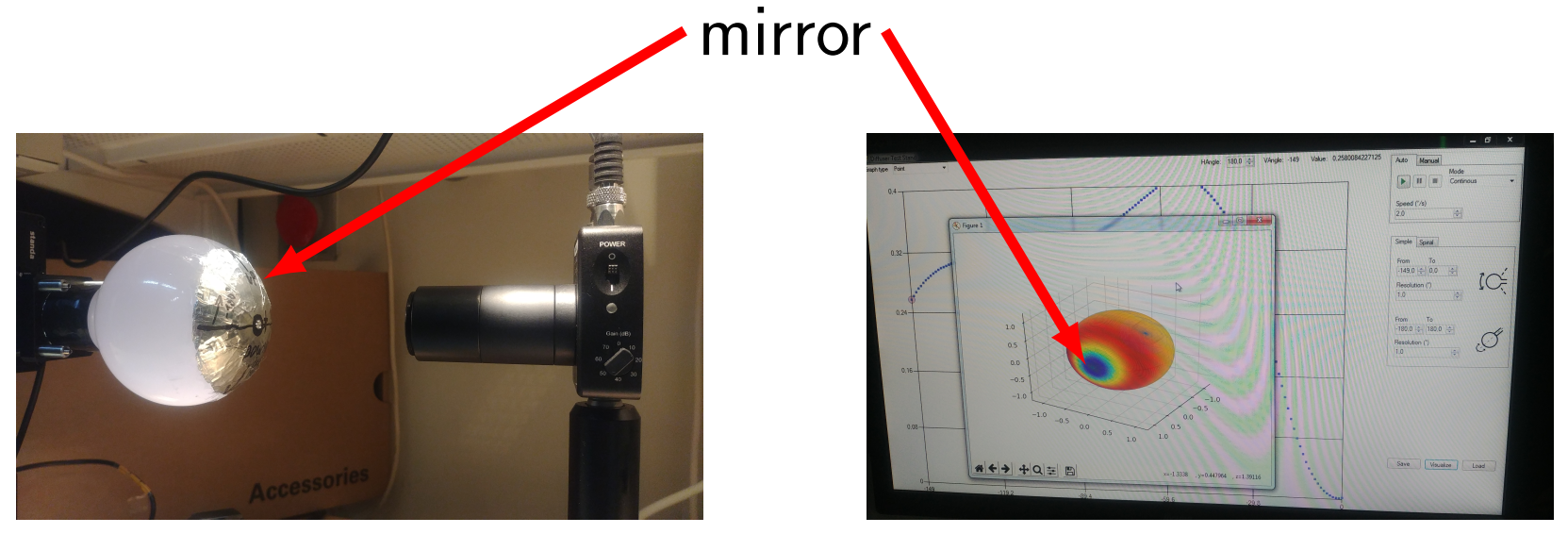}
	\vspace*{-4ex}
	\caption{Far scan result. The test was performed with a mirror on the top to improve the isotropy in a backward region. Right: The scanner control interface with scan visualization.}
	\label{fig:white_diffuser_scan_bench}
	\vspace*{-2ex}
\end{figure}

During the research works on the test bench, two scan methods were developed: the so-called "near scan" (Fig. \ref{fig:near_scan}), also called differential scanning, and "far scan" (Fig. \ref{fig:white_diffuser_scan_bench}), also called integral scanning.
Near scan allows the examination of a small structural changes in optical elements, mainly in diffusers. It explores the emission of light in a very narrow angle from a point lying on the surface of the tested element. This scan allows, inter alia, to determine the impact of doping of the diffuser material to its attenuation. 

The far scan allows us to determine the actual light distribution at any distance from the diffuser (or other tested element). In the scan configuration, the detector "observes" the entire visible surface of the tested element. Such a scan already taking into account the absorption and scattering coefficients of a medium in which the measurement is performed. The data collected from the far scan performed from a distance of 14 cm (determined experimentally in relation to the diffuser size) were used as an input data for the simulation of light propagation in water.

\begin{figure}[H]
	\vspace*{-2ex}
	\centering
	\includegraphics[width=0.8\textwidth,keepaspectratio]{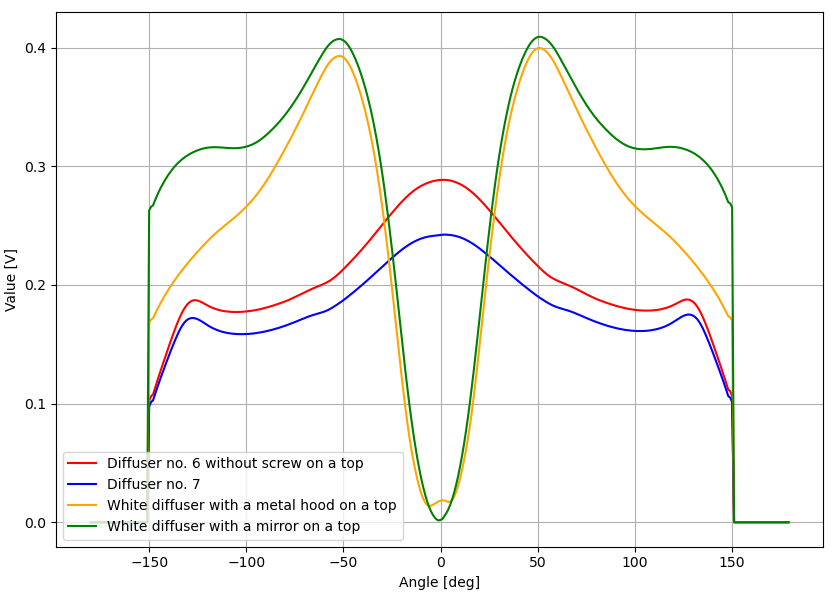}
	 \vspace*{-2ex}
	\caption{Scans results comparison.}
	\label{fig:scan_results_comparison}
	\vspace*{-2ex}
\end{figure}

\begin{figure}[H]
	\vspace*{-2ex}
	\centering
	\includegraphics[width=0.8\textwidth,keepaspectratio]{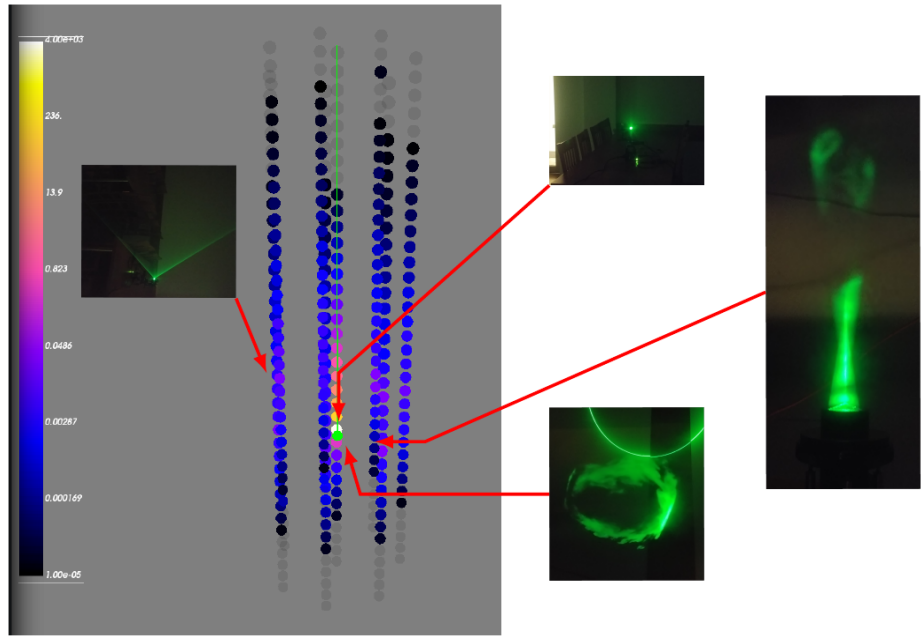}
\caption{Simulation result for the laser attached to the center string of a cluster, equipped with Axicon (\ang{20}), for upgoing beam in water. The pictures on the right shows the real view of the laser beam, shaped into a cone in water mist (\SI{65}{\percent RH}), seen from the indicated points of the simulated cluster.}
	\label{fig:real_axicon_effect}
	 \vspace{-2ex}
\end{figure}

An Initial research was carried out on a set of three different diffusers used in the Baikal-GVD experiment till the end of 2020. 
The research revealed differences in the light distribution and in the attenuation of individual diffuser materials (Fig. \ref{fig:scan_results_comparison}). A number of tests were also carried out to improve the light distribution to illuminate the Optical Modules below the laser horizon in the Baikal-GVD telescope.
One of the tested solutions was the propagation of light in the form of a cone. For this purpose, after changing the configuration of the test bench and equipping it with a dedicated optical components, a test scan was performed and the data was provided in the simulation. An exemplary visualization of the conical emission is shown in the Fig. \ref{fig:real_axicon_effect}.


\pagebreak
\thispagestyle{empty}
\acknowledgments
This research is supported by the Plenipotentiary Representative of the Government of the Republic of Poland at JINR in Dubna under the Projects No. 75/09/2020 and by the Ministry of science and higher education of the Russian Federation under the contract No. 075-15-2020-778.
We would also like to express our gratitude to our colleagues from the Polish team, A. Bhatt, D. Góra, J. Stasielak and M. Wiśniewski, who contributed to this article.

\end{document}